# Magnetic doping effects on the superconductivity of Y$_{1-x}$M$_x$Ba$_2$Cu$_3$O$_{7-\delta}$ (M = Fe, Co, Ni)

Hamideh Shakeripour[a]*, Seyed Sajjad Hosseini[a], Seyedeh Sara Ghotb[a,b], Behnaz Hadi-Sichani[a],

Sepideh Pourasad[a]

[a] Department of Physics, Isfahan University of Technology, Isfahan 84156-83111, Iran
[b] Département de physique & RQMP, Université de Sherbrooke, Sherbrooke, Québec J1K 2R1, Canada
*Corresponding author:

Hamideh Shakeripour (hshakeri@iut.ac.ir)

Phone: +983133913722, Fax: +983133912376

**Abstract**

The discovery of superconductivity in copper oxide compounds has attracted considerable attention over the past three decades. The high transition temperature ($T_c$) in these compounds, exhibiting proximity to an antiferromagnetic order in their phase diagrams, remains one of the main areas of research. It is believed that magnetic fluctuations provide substance for the exotic superconductivity observed in these compounds. The present study attempts to introduce Fe, Co and Ni magnetic impurities into the superconducting cuprate YBa$_2$Cu$_3$O$_{7-\delta}$ with the aim of exploring the $T_c$ behavior. The solid-state synthesis method is exploited to prepare fully oxygenated Y$_{1-x}$M$_x$Ba$_2$Cu$_3$O$_{7-\delta}$ (Y$_{1-x}$M$_x$-123) (M = Co, Fe, Ni) samples with low levels of doping (0.00000 ≤ x ≤ 0.03000). Systematic measurements are then employed to assess the synthesized samples using AC magnetic susceptibility, electrical resistivity and X-ray diffraction (XRD). The measurements revealed an increase in $T_c$ as a result of magnetic substitution for Y. However, the study of non-magnetic dopings on the fully oxygenated Y$_{1-x}$M'$_x$Ba$_2$Cu$_3$O$_{7-\delta}$ (Y$_{1-x}$M'$_x$-123) (M' = Ca, Sr) samples showed a decrease in $T_c$. Quantitative XRD analysis further suggested that the internal pressure could have minor effects on the increase in $T_c$. The normal state resistivity vs temperature showed a linear profile, confirming that the samples are at an optimal doping of the carrier concentration.

# 1. Introduction

Metal oxide superconductors have emerged as a leading subject of research due to some promising properties that could revolutionize the industrial and environmental aspects of the modern technology. In this respect, the high-$T_c$ cuprate superconducting systems of $RBa_2Cu_3O_{7-\delta}$ (R = rare earth ions or Y, except for Tb, Ce, Pm and Pr) are the most demanded due to a number of reasons. The most preferable feature of these materials is probably found in their relatively high $T_c$, which is above the liquid nitrogen temperature. Secondly, different well-established synthesis techniques have been successfully designed to fabricate polycrystalline and single crystalline samples with large sizes from these oxides. Additionally, the physical properties of these materials could be feasibly tuned by partial or complete atomic substitutions [1-5].

Various studies have been conducted on atomic substitutions in $YBa_2Cu_3O_{7-\delta}$ and the subsequent effects on $T_c$ have been addressed. Substitutions in YBCO would generally affect the carrier concentration, thus leading to significant changes in the superconducting properties. Substitution of Y by monovalent, divalent and trivalent elements [6-8] has revealed that while no impurity phases are observed in $Y_{1-x}A_xBa_2Cu_3O_{7-\delta}$ (A = K, Rb, Cs) and $Y_{0.9}M_{0.1}Ba_2Cu_3O_{7-\delta}$ (M = Mg, Ca, Sr, Ba), the resulting $T_c$ values are decreased [8]. For instance, Ca substitution leads to a decrease of $T_c$ in the fully oxygenated samples in spite of the fact that it has an ionic radius close to Y. However, it is intuitively established that no $T_c$ variation takes places by doping elements with ionic radii similar to Y [8]. As an example of trivalent element dopings, Al [9] would decrease $T_c$, which is reportedly due to the gradual changes occurring in the YBCO crystal lattice [9]. All

RBa$_2$Cu$_3$O$_{7-\delta}$ systems exhibit a $T_c$ of approximately 90 K [10], leading to the assumption that the R site is electronically isolated from the conduction electrons. However, in the fully oxygenated Gd$_{1-x}$Y$_x$Ba$_2$Cu$_3$O$_7$ samples, it was found that the $T_c$ degradation could occur by increasing x, which was suggested to be mainly related to the effects of the R elements [11].

Concerning the replacement of elements for the Cu site [12-17], studies have shown that both magnetic and diamagnetic ions would radically decrease $T_c$ in a similar manner [10, 18]. It was observed that substitution by transition-metal ions resulted in the $T_c$ reduction or disappearance of superconductivity [19]. In addition, $T_c$ is reported to be strongly correlated with the magnitude of the paramagnetic moments. In the YBa$_2$Co$_x$Cu$_{3-x}$O$_{7-y}$ system, Co substitution leads to a transition from an orthorhombic structure to a tetragonal one for x when it is between 0.05 and 0.1, while $T_c$ would further decrease and render the compound non-superconducting by reaching x = 0.4 [12, 14]; however, the suppressed $T_c$ can be improved by the substitution of Ca for Y. This effect could be explained on the basis of the increased hole concentration and bond length variations [15]. In YBa$_2$Cu$_{3-x}$M$_x$O$_{7-\delta}$ (M = Zn, Ni) samples, $T_c$ was depressed even more quickly in the non-magnetic Zn-doped compound than the Ni-doped one, lending support to the conclusion that superconducting properties in some materials are much more sensitive to local structural disorders than magnetic interactions.

The present study is concerned with introducing a *very low level* of *magnetic* impurities *for the Y site* into the YBa$_2$Cu$_3$O$_{7-\delta}$ system and exploring the subsequent $T_c$ variations. To the best of our knowledge, there have not been any systematic studies on the substitution of the magnetic elements for Y and the response of the cuprates' $T_c$ at these *very low levels* of concentration. Magnetic susceptibility and electrical resistivity measurements are accordingly exploited to assess the transitions to the superconducting state. It is found that the substitution of a certain quantity of Co,

Fe and Ni for Y leads to the increase of $T_c$, while the substitution of two non-magnetic elements, i.e. Ca and Sr, results in the $T_c$ degradation. The crystal structure analysis also shows that chemical pressure might have minor effects on the observed enhancement in $T_c$. According to these findings, the increase in $T_c$ through magnetic doping confirms the effective role of magnetically-mediated pairing in cuprates.

## 2. Materials and methods

### 2.1. Method

Polycrystalline fully oxygenated samples of the series $Y_{1-x}M_x$-123 (M = Co, Fe, Ni) and $Y_{1-x}M'_x$-123 (M' = Ca, Sr) were prepared via the solid-state reaction method using high-purity $BaCO_3$, $CaCO_3$, $SrCO_3$, $Y_2O_3$, CuO, $Co_3O_4$ and $Fe_2O_3$ powders (99.99%). The stoichiometric proportions of the powders were weighed ($10^{-5}$ g precision), mixed together, ground and pressed into pellets to be calcined in alumina crucibles. The sintering process was carried out under a pure and dry oxygen flow to get the desired full oxygen content [20]. To have a systematic comparison of the results, all samples were prepared under identical conditions, sample preparation and heating process [20], and sintered together, leading to excellent reproducibility of the results. The superconducting samples were prepared with x = 0.00000, 0.00100, 0.00200, 0.00300, 0.00400, 0.00500, 0.00600, 0.00700, 0.00800, 0.00900, 0.01000, 0.02000, 0.03000, 0.04000, 0.05000, 0.06000 and 0.07000.

### 2.2. Characterizations and measurements

Resistivity was measured using the standard DC four-probe technique. AC magnetic susceptibility was measured using a Lake-Shore 7000 AC susceptometer in the magnetic field of 0.8 A/m and at a frequency of 333 Hz. The transition temperature, $T_c$, is defined as the temperature below which the real part of susceptibility deviates from its normal-state behavior and drops to the diamagnetic superconducting phase. The XRD measurements were used to examine the phase purity and to obtain lattice parameters using a Philips XRD instrument and Cu-K$_\alpha$ radiation. The X-ray patterns were analyzed in a systematic manner using the Rietveld refinement by the Fullprof Suite software. The oxygen contents of the samples (7-δ) were obtained using the data related to the refined patterns [20]. It was shown that the oxygen content of all doped samples was effectively constant with respect to the dopant concentrations. In addition, the iodometric titration method was used to determine the oxygen contents of the samples [21]. To put it briefly, the results of both oxygen determination methods employed were consistent, showing an optimum and constant value of 6.97±0.02, independent of the degree of the substitutions.

## 3. Results and discussion

### 3.1. AC susceptibility

The AC magnetic susceptibility results of the series $Y_{1-x}M_x$-123 (M = Co, Fe, Ni) are depicted in **Fig. 1**. It is found that $T_c$ would initially increase with magnetic element dopings up to the optimal contents, x ≈ 0.01, 0.004, 0.004 for Co, Fe and Ni dopants, respectively. While the pure sample has a $T_c$ = 90.91 K, it is observed that the Co (Fe) impurity of nearly 1% (0.4%) would increase it to 92.75 K (**Fig. 1(a,b)**). Ni substitution also shows the same trend (**Fig. 1(c)**). The inset of **Fig. 1(a)** shows the $T_c$ variations as a function of the Co content. $T_c$ was raised to a maximum at the optimal value of the Co dopant, before it was declined due to excessive doping. The Co content dependence formed a dome-shaped curve. The same trend was found for Fe and Ni doped samples as well.

The observed increasing rate of $T_c$ by magnetic substitutions is found to be in contradiction to the Abrikosov-Gorkov theory, which predicted a reduction in $T_c$ [22, 23]. However, the same study of Ca and Sr dopings (non-magnetic impurities) reveals no increase in $T_c$ (**Fig. 2**). Moreover, introducing Fe (0.4%) and Co (1%) led to equal increases in $T_c$. Also, compared to the nearly 0.4% Co-doped sample, the 0.4% Fe-doped sample exhibited the higher $T_c$ of 0.8 K in spite of their nearly equal ionic radii, while Fe holds a higher intrinsic magnetic moment, as compared to Co. However, no increase in $T_c$ is found in Ca- and Sr-doped compounds, which might suggest the effective role of magnetic atoms in enhancing $T_c$. Eventually, once the magnetic element concentration is increased beyond the optimum value, $T_c$ starts to decline (**Fig. 3**). Overdoping of the hole concentration and the presence of large magnetic moments of impurities (above the optimum value) could lead to the reduction of $T_c$.

### 3.2. Electrical resistivity

**Fig. 4** shows $\rho(T)$ measurements for the as-synthesized samples. Clearly, the optimum doping in each system resulted in the $T_c$ enhancement, which is in agreement with the susceptibility data (**Fig. 4(a,c,e)**). Moreover, all samples exhibited a linear temperature dependence, as shown in Eq. (1), over the whole temperature range (see **Fig. 5**, for instance) in the normal state down to above $T_c$.

$$\rho(T) = \rho_0 + AT \tag{1}$$

The linear behavior of $\rho(T)$, a distinctive feature observed at optimal dopings in high-$T_c$ cuprates which present no depletions related to the pseudogap [24], shows that all samples are fully oxygenated at the optimal carrier concentration [24]. While the origin of the linear $\rho(T)$ in cuprates is under debate, it is attributed to the antiferromagnetic (AFM) spin fluctuation scatterings in organic [25] and heavy-fermion [26] superconductors, where the magnetically mediated superconductivity has been established [26-33].

Furthermore, the slope of resistivity vs temperature ($A$), which is related to the carrier concentration [34], does not change by magnetic doping (**Fig. 5**), thereby implying another possible factor (perhaps dopant spins) influencing the $T_c$ behavior. There is a nearly small decrease in $A$ up to the optimum dopings for Co and Ni-containing samples (**Fig. 4(a,e)** or **Fig. 6(a,b)**). It could be, therefore, inferred that the substitution of $Co^{2+,3+}$ and $Ni^{2+}$ for $Y^{3+}$ might lead to a slight increase in the hole concentration, or the magnetic dopants would cause changes to the carrier concentration. The latter may not be explained by simple chemistry; this is because, in spite of other atoms in the system, Fe, Co and Ni possess local moments. However, the clear observation of the constant $A$ in all Fe concentrations (**Fig. 5**) confirms the former.

### 3.3. XRD and internal pressure effects

The XRD measurements revealed that all samples are single-phased, showing an orthorhombic crystal structure of identical perovskite-type with *Pmmm* symmetry; they were matched with the standard ICDD-JCPDS 38-1433 pattern. The Rietveld refinement method was used to obtain the crystallographic data of the samples (**Fig. 7(a)**). The normalized XRD patterns, indicating the phase formation of the samples for different Co, Fe and Ni dopants, are shown in **Fig**. **7(b-d)**. The internal pressure effect might be of interest in these compounds as a factor of $T_c$ variations. Internal pressure can be compared with the effects of the externally applied pressure [35]. Efforts were made in this study to simulate the effect of the chemical pressure on $T_c$ through doping. It is worth mentioning that variations of $T_c$ under pressure have shown a striking diversity for different high-$T_c$ superconductors due to their highly anisotropic layered structure, such that no clear picture of this phenomenon has emerged yet. As reported in the earlier studies, the applied pressure on the YBa$_2$Cu$_3$O$_7$ compound [36] could only have a slight effect on $T_c$, which is in sharp contrast to what has been observed for La, Tl, and Hg-based cuprates. Application of 2.0 GPa pressure increased $T_c$ by only 1.3 K [36, 37], which was also reported to be 0.03 K/kbar [38] or 0.7 K for 16 kbar [11]. However, a 1.8 K $T_c$ enhancement was observed in this study for the Co (Fe) doping of only 2% (0.4%), the origin of which might not be ascribed to the internal pressure only; rather, another factor (such as impurity spins) might play a role (as implied by the analogous 1.25 K increase in $T_c$ for 0.4% Ni doping). Furthermore, the rate of increase in $T_c$ vs pressure has been reportedly too small and linear [11, 36, 37], which is inconsistent with the dome-shaped behavior observed in **Fig. 3(a,b)**. Additionally, the decline of $T_c$ upon increasing x in Gd$_{1-x}$Y$_x$Ba$_2$Cu$_3$O$_7$ [11] samples was attributed to the chemical pressure effect, whereas the obtained results in the present study suggest that the reduction of the Gd content (hence, its depressed magnetic effect) might be the reason behind the decrease of $T_c$.

Other studies [39, 40] have reported the effects of uniaxial pressure application on $T_c$ for the YBa$_2$Cu$_3$O$_7$ system. Drawing upon their results, Lin *et al*. [41] exploited the changes incurred by uniaxial pressure applications in the lattice parameters ($a$, $b$, $c$), proposing an empirical relation for calculating the corresponding $T_c$ values. The present study utilizes this relation to calculate $T_c$ values (called $T_{c,calc}$) based on the lattice parameters determined from the XRD analysis. For the entire series of R-123 systems, the $T_c$ equation is given by:

$$T_c - T_c^r = \frac{\partial T_c}{\partial \epsilon_1}\frac{a - a_0}{a} + \frac{\partial T_c}{\partial \epsilon_2}\frac{b - b_0}{b} + \frac{\partial T_c}{\partial \epsilon_3}\frac{c - c_0}{c} \qquad (2)$$

, where $T_c^r$, a$_0$, b$_0$ and c$_0$ are the $T_c$ and lattice parameters for a reference material (pure Y-123 system) and $\epsilon_i$ denotes the strains along the $i=1,2,3$ directions. Use was made in this process of the measured values of d$T_c$/d$P_j$ (where $P_j$ is the uniaxial pressure along the $j = a, b, c$ crystal axes) and the derived values of $\partial T_c/\partial \epsilon_i$ ($i=1,2,3$) (data from [39, 40]). The results are shown in **Table 1** for the three magnetic dopings of Co, Fe and Ni. According to Eq. (2), there is no agreement between the calculated and measured values of $T_c$, thereby approving that the behavior of the exotic experimental $T_c$ values ($T_{c,exp}$) is independent of the pressure notion.

### 3.4. SEM

The scanning electron microscopic (SEM) images of the optimally doped samples (with $Co_{0.02}$, $Fe_{0.004}$ and $Ni_{0.004}$) are shown in **Fig. 8**. It could be clearly seen that the porosity level is very low. The grains with different sizes are distributed in the samples. It is observed that the large grain sizes are present in the samples (around 20 μm), which is consistent with the observed maximum activation energy and the maximum measured and estimated $J_c$ [42] for the optimally doped samples.

### 3.5. Comparison with the previous reports

Introduction of magnetic impurities to the YBCO system has been a subject of interest among researchers. It has now been well established that the Cu atom is one of the key elements in the cuprate family, so that even its partial substitution (either magnetic or non-magnetic) leads to the reduced superconductivity. Maeno *et al.* [13] substituted large quantities of Co, Fe and Ni (x = 0.033) for at least 3.3% of magnetic impurities at the Cu site in $YBa_2(Cu_{1-x}M_x)_3O_{7-\delta}$, observing the reduction of $T_c$. It could be deduced that by this substitution, the spin of magnetic dopings could break copper pairs and reduce $T_c$. However, very low amounts of magnetic moment dopings in the present study contributed to the appearance of superconductivity at slightly higher temperatures. Furthermore, the XRD analysis revealed no shifts in any of the central peaks of the XRD patterns in all $Y_{1-x}M_x$-123 samples, when compared with those of the pure material (see **Fig. 7**), which may be interpreted as the substitution of the dopant for the Y atom.

Based on the previous studies [10, 12-16], magnetic dopings in the Cu site led to the decrease rather than increase of $T_c$ in the samples. According to the general consensus on $T_c$ suppression by magnetic factors, the 0.4% Fe-doped sample is found to have an exotically higher $T_c$ than the 0.4% Co-doped sample (nearly 0.8 K higher), in spite of the fact that Fe has a higher magnetic moment and should, therefore, suppress superconductivity more strongly.

Another point worth investigating is whether the pristine sample with $T_c \approx 91$ K is slightly off the optimally doped region, so that $T_c$ is maximized by impurity doping. This effect is found in the results reported by Suard *et al*. [15], where the suppressed $T_c$ in the low $T_c$ Y-123 samples was improved by Ca substitution for Y due to the increase of the hole concentration and bond length variations. However, adding analogous small amounts of Ca or Sr in the present study reduced $T_c$ strongly (in contrast to the magnetic dopings), which means that the pristine sample has been optimally doped. Moreover, considering the temperature dependence of the normal state resistivity of YBCO, as reported in the literature, the $\rho$(T) behavior would not be linear in the normal state over the whole temperature range if the pure sample had a low carrier concentration or was slightly off the optimally doped region. Therefore, the rise in $T_c$ due to the magnetic doping is a new observation which needs to be explained.

## 4. Conclusions

To summarize, introducing magnetic and non-magnetic elements as partial substitutions for Y in the $YBa_2Cu_3O_{7-\delta}$ system suggested the possible spin fluctuation effect as the key mechanism of pairing in cuprates. The increase of $T_c$ by magnetic element dopings was confirmed by AC magnetic susceptibility and resistivity measurements in the fully oxygenated $Y_{1-x}M_x$-123 (M = Co, Fe, Ni) samples. $T_c$ degradation was observed upon non-magnetic element dopings (Ca, Sr) in the same compound. XRD analysis also suggested that the internal pressure could have a minor effect on the increase in $T_c$. The slope of the normal state resistivity vs temperature is nearly constant for various magnetic doping concentrations. Thus, it could be deduced that the increase in $T_c$ is mainly related to the spin of the dopants rather than any internal pressure effects or variations of the carrier concentration.

**Acknowledgements**

Authors would like to thank M. Abbasi and Kh. Rahmani for their technical assistance. They also express their gratitude to Professors H. Salamati, M. Akhavan, P. Kameli, H. Ahmadvand, J. Paglione, R. Greene, K. Behnia, L. Taillefer and F. Davar for insightful discussions. We also acknowledge support from the INSF.

**Table 1.** The lattice parameters obtained from XRD measurements in $Y_{1-x}M_x$-123 (M = Co, Fe, Ni) samples and the experimental ($T_{c,exp}$) and calculated ($T_{c,calc}$) values of $T_c$. The $T_{c,exp}$ values are extracted from **Figs. 1** and **3**.

| x | a (Å) | b (Å) | c (Å) | $T_{c,exp}$ (K) | $T_{c,calc}$ (K) |
|---|---|---|---|---|---|
| 0.000 | 3.8249 | 3.8827 | 11.6835 | 90.91 | 90.91 |
| **$Y_{1-x}Co_x$-123** | | | | | |
| 0.002 | 3.8237 | 3.8818 | 11.6895 | 91.85 | 89.37 |
| 0.005 | 3.8222 | 3.8810 | 11.6964 | 92.21 | 89.34 |
| 0.010 | 3.9208 | 3.8801 | 11.6995 | 92.75 | 89.31 |
| 0.020 | 3.8201 | 9.8796 | 11.7018 | 92.89 | 89.29 |
| 0.030 | 3.8106 | 3.8791 | 11.7009 | 92.11 | 88.74 |
| **$Y_{1-x}Fe_x$-123** | | | | | |
| 0.002 | 3.8221 | 3.8821 | 11.6929 | 92.26 | 90.79 |
| 0.003 | 3.8225 | 3.8819 | 11.6984 | 92.28 | 90.83 |
| 0.004 | 3.8215 | 3.8816 | 11.72 | 92.72 | 90.79 |
| 0.010 | 3.8235 | 3.8828 | 11.6935 | 92.24 | 90.83 |
| **$Y_{1-x}Ni_x$-123** | | | | | |
| 0.002 | 3.8201 | 3.8853 | 11.6871 | 91.67 | 90.48 |
| 0.004 | 3.8215 | 3.8921 | 11.7064 | 92.16 | 90.21 |
| 0.006 | 3.8215 | 3.8852 | 11.6937 | 91.93 | 90.58 |
| 0.020 | 3.8196 | 3.8817 | 11.6758 | 89.81 | 90.63 |

**Captions to illustrations:**

**Fig. 1** AC magnetic susceptibility data of (a) $Y_{1-x}Co_x$-123, (b) $Y_{1-x}Fe_x$-123 and (c) $Y_{1-x}Ni_x$-123 samples in the rising $T_c$ region. The arrow shows that $T_c$ improves with increasing the magnetic dopant to an optimum value of (a) $x \approx 0.01$ for Co, (b), (c) $x \approx 0.004$ for Fe or Ni dopant.

**Fig. 2** AC magnetic susceptibility results of non-magnetic dopants in (a) $Y_{1-x}Ca_x$-123 and (b) $Y_{1-x}Sr_x$-123 samples. The arrow shows that $T_c$ is decreased with a low magnitude of Ca and Sr dopant enhancement.

**Fig. 3** AC magnetic susceptibility data of (a) $Y_{1-x}Co_x$-123, (b) $Y_{1-x}Fe_x$-123 and (c) $Y_{1-x}Ni_x$-123 samples in the falling $T_c$ region. The arrow shows that $T_c$ is declined when the magnetic dopant is added beyond the optimum level ($x \approx 0.01, 0.004, 0.004$ for Co, Fe and Ni dopants, respectively). (Insets) $T_c$ vs Co, Fe, Ni content. The values of $T_c(x)$ are extracted from **Fig. 1(a,b,c)** and **3(a,b,c)**. Error bars come from re-making samples and extracting procedure of $T_c$. The initial rise of $T_c(x)$ with adding the magnetic impurity (the left side of the dome curve) is surprising, challenging our understanding.

**Fig. 4** Electrical resistivity measurements of (a,b) $Y_{1-x}Co_x$-123, (c,d) $Y_{1-x}Fe_x$-123 and (e,f) $Y_{1-x}Ni_x$-123 samples in the rising (a,c,e) and falling (b,d,f) $T_c$ regions. *Left*: in compliance with **Fig. 1**, $T_c$ increases up to an optimum doping level. *Right*: $T_c$ decreases when the magnetic dopant goes beyond the optimum level.

**Fig. 5** Selected electrical resistivity measurements of $Y_{1-x}Fe_xBa_2Cu_3O_{7-\delta}$ in (a) rising and (b) falling $T_c$ regions. $\rho(T)$ vs $T$ for all magnitudes of doping is linear, $\rho(T) = \rho_0 + AT$, from room temperature down to near above $T_c$ (the same trend is seen for all Co- and Ni-doped samples). (*Top*): Low doped, (*Bottom*): overdoping of the Fe dopant.

**Fig. 6** The electrical resistivity measurements of (a) $Y_{1-x}Co_xBa_2Cu_3O_{7-\delta}$ (b) $Y_{1-x}Ni_xBa_2Cu_3O_{7-\delta}$. $\rho(T)$ vs $T$ for all magnitudes of doping is linear, $\rho(T) = \rho_0 + AT$, from room temperature down to near above $T_c$. There is a nearly small decrease in $A$ up to the optimum dopings for Co and Ni-containing samples. It could be inferred that the substitution of $Co^{2+,3+}$ and $Ni^{2+}$ for $Y^{3+}$ might lead to a slight increase in the hole concentration.

**Fig. 7** (a) Analysis of the XRD data by Rietveld refinement for $Y_{0.98}Co_{0.02}$-123 (for instance); XRD patterns of (b) $Y_{1-x}Co_x$-123, (c) $Y_{1-x}Fe_x$-123 and (d) $Y_{1-x}Ni_x$-123 samples.

**Fig. 8** SEM micrographs of (a) $Y_{0.98}Co_{0.02}$-123, (b) $Y_{0.996}Fe_{0.004}$-123 and (c) $Y_{0.996}Ni_{0.004}$-123 samples. It is seen that the grain sizes are around 20 µm. Also, porosity is very low and grains coupling is high. The micrographs are shown in 5 µm resolutions.

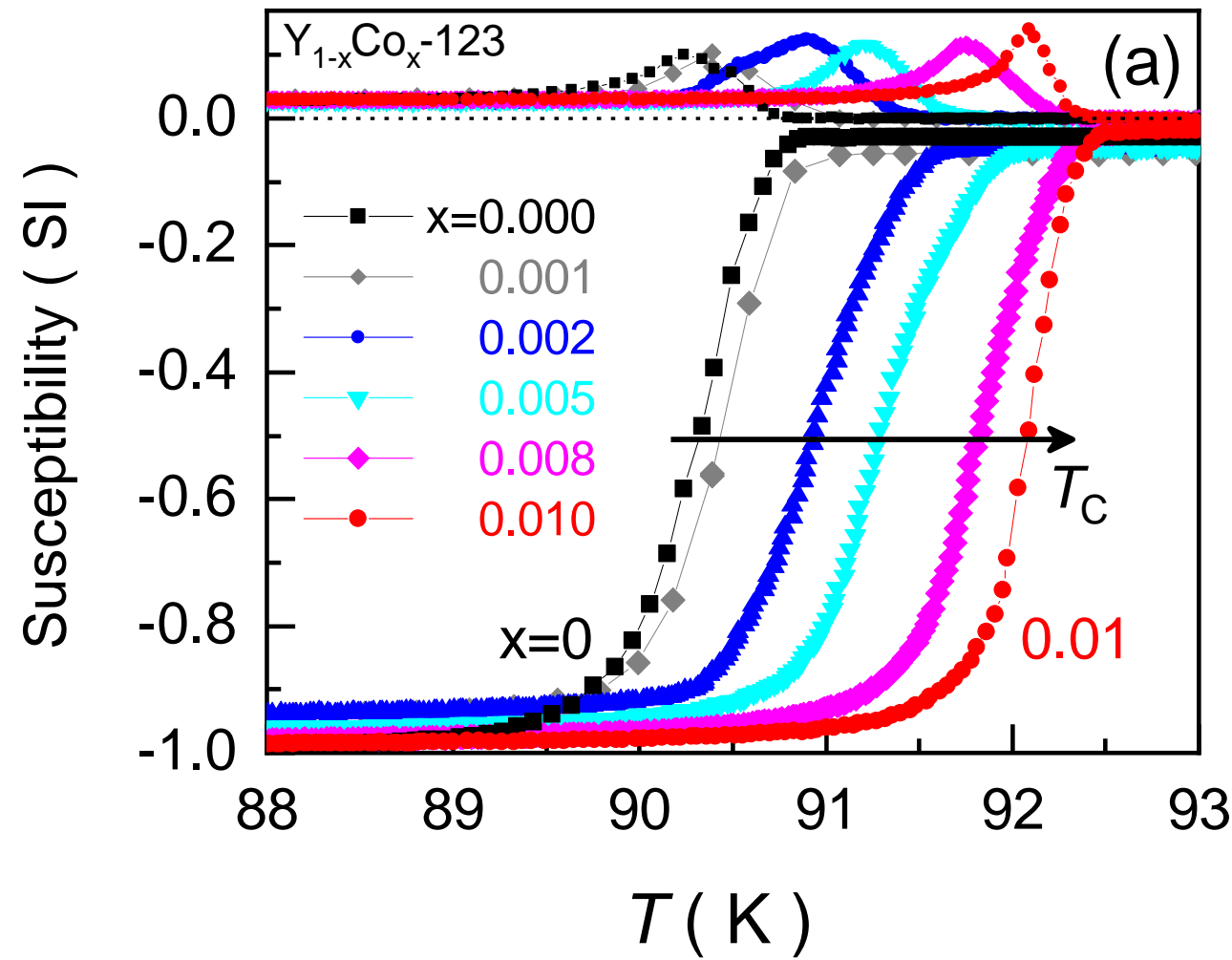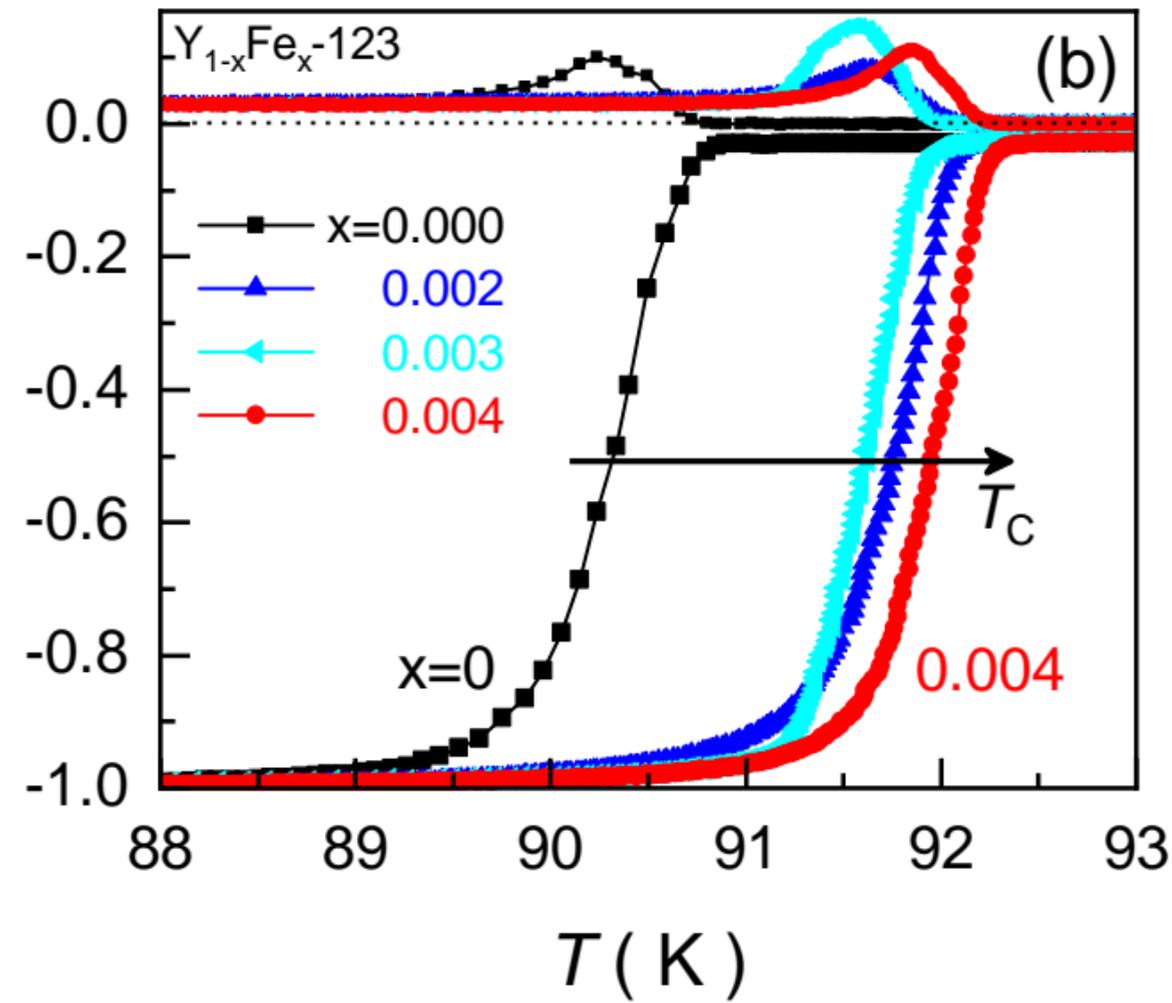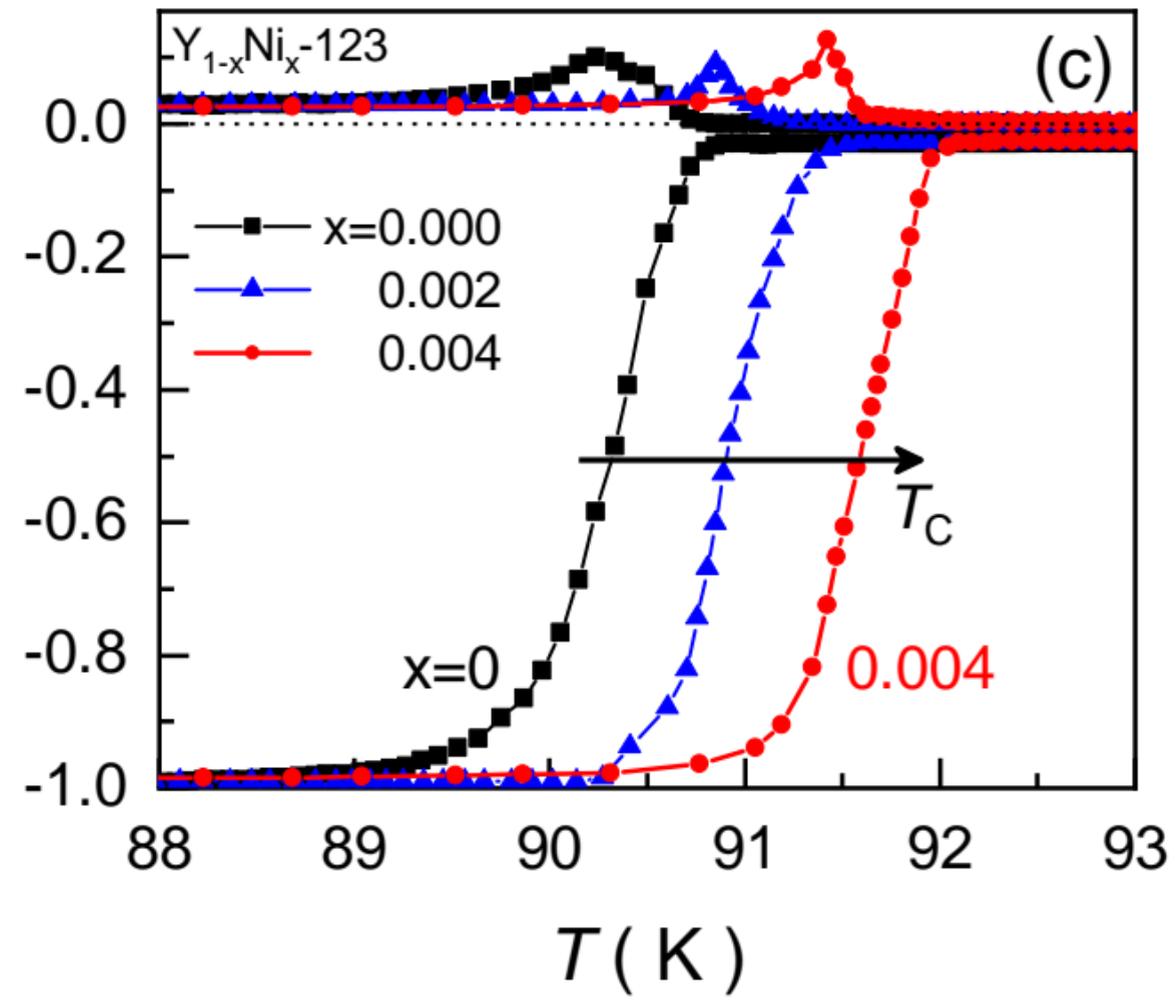

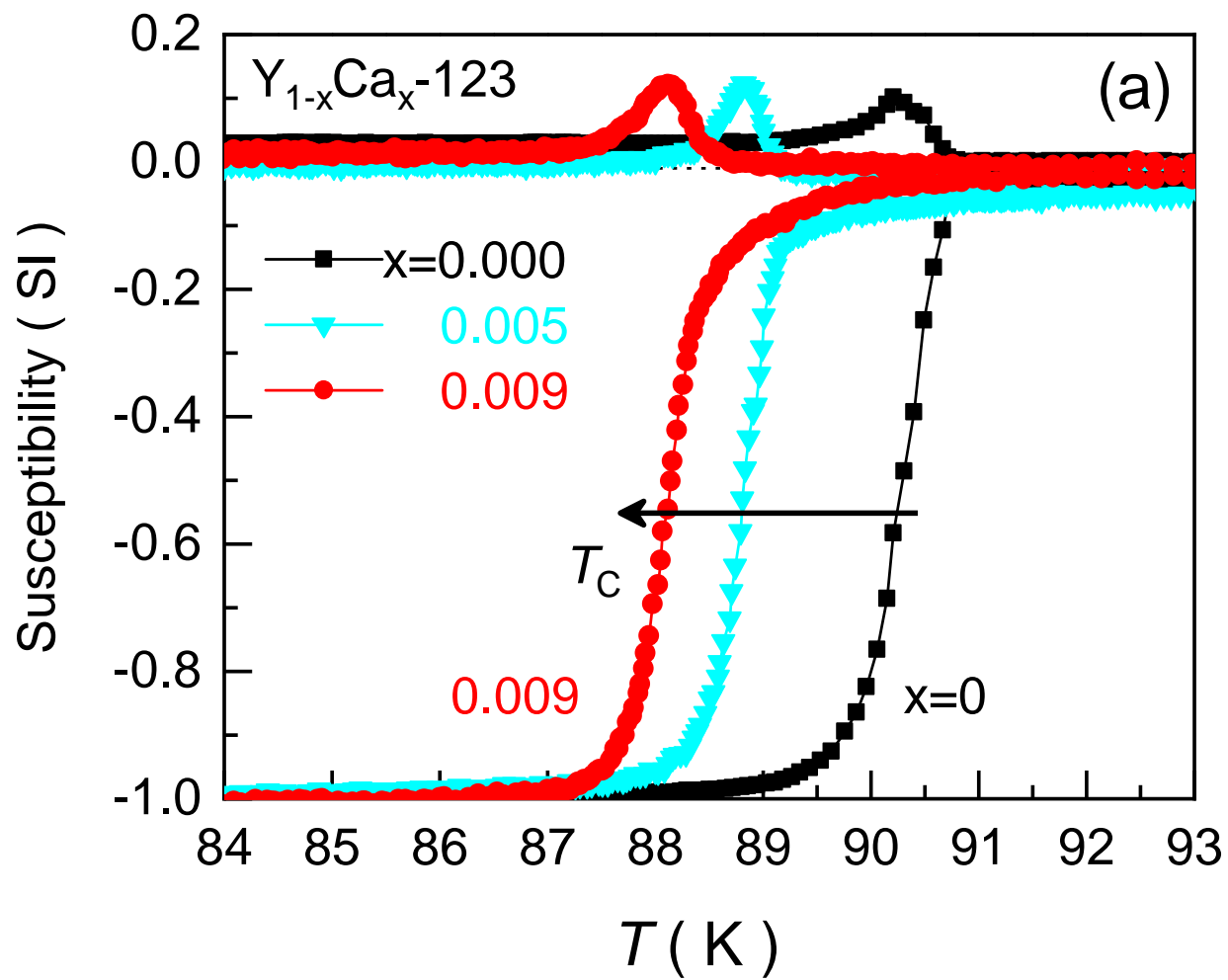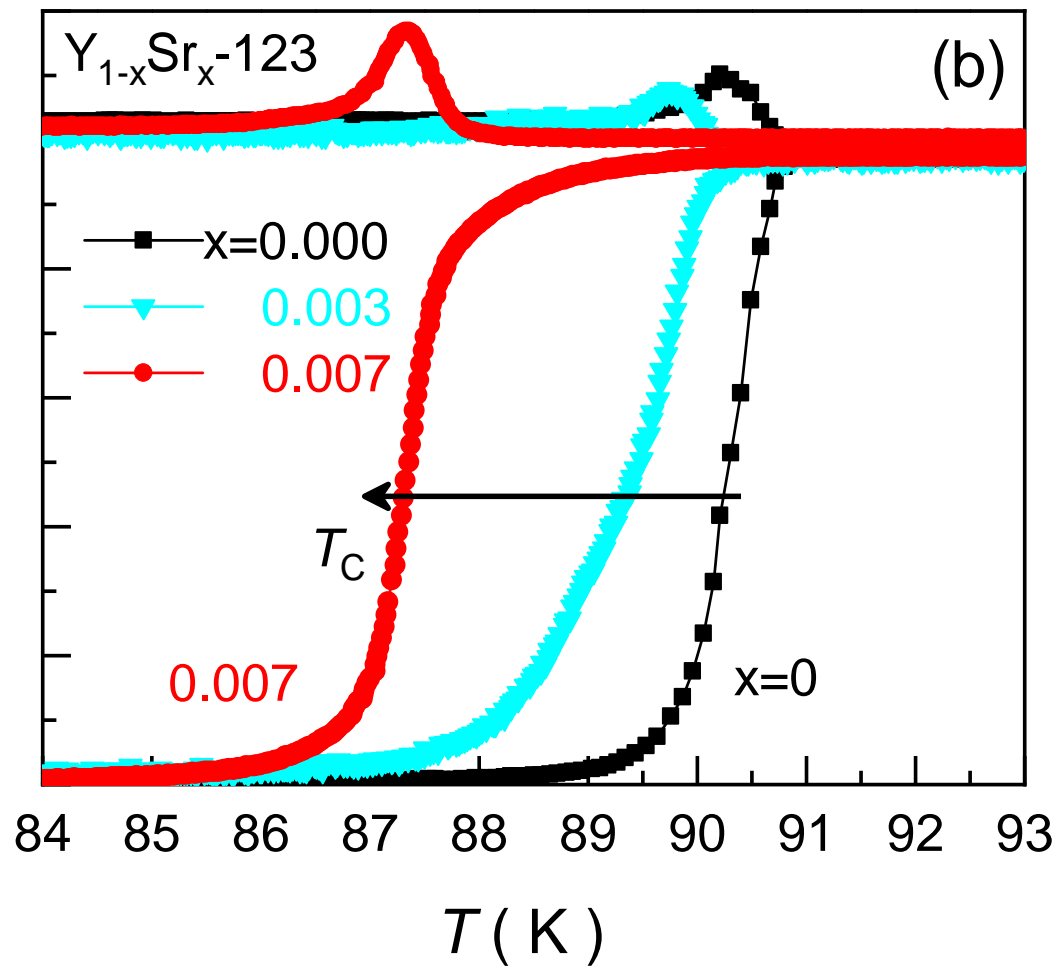

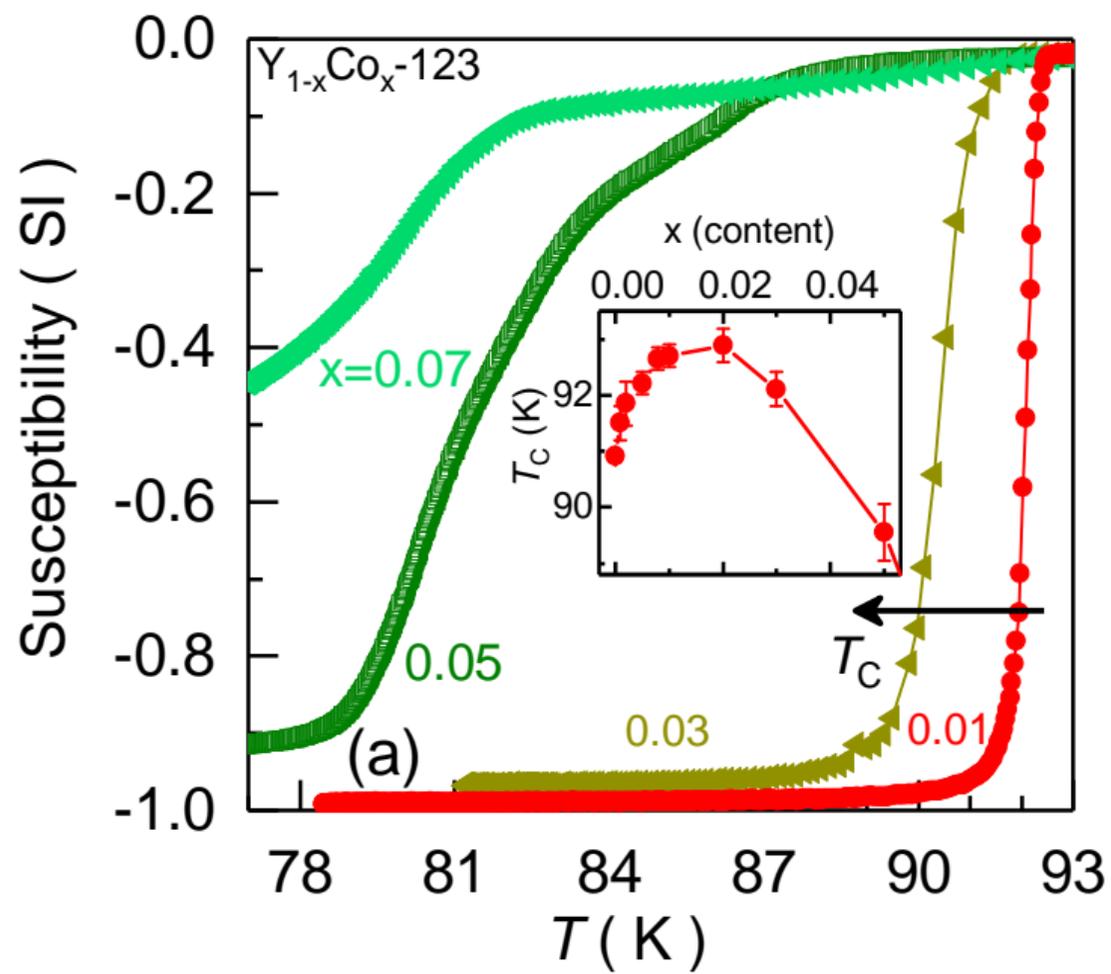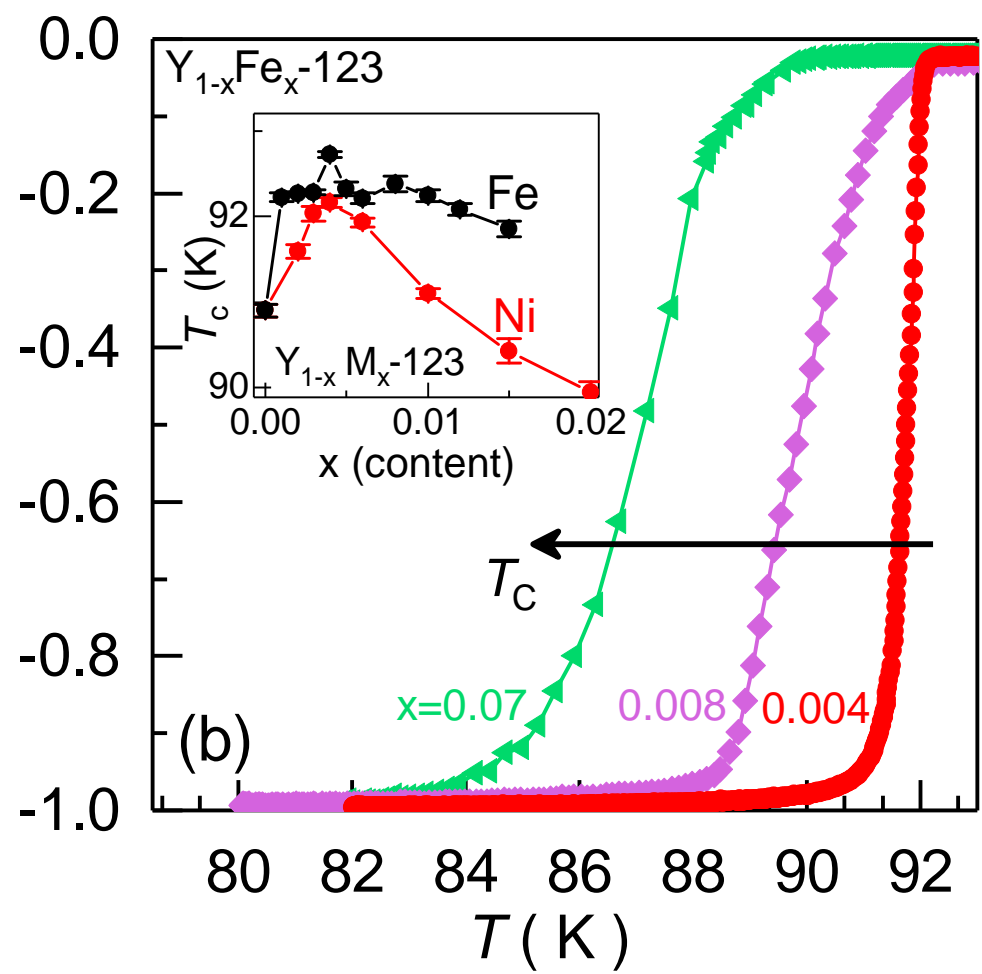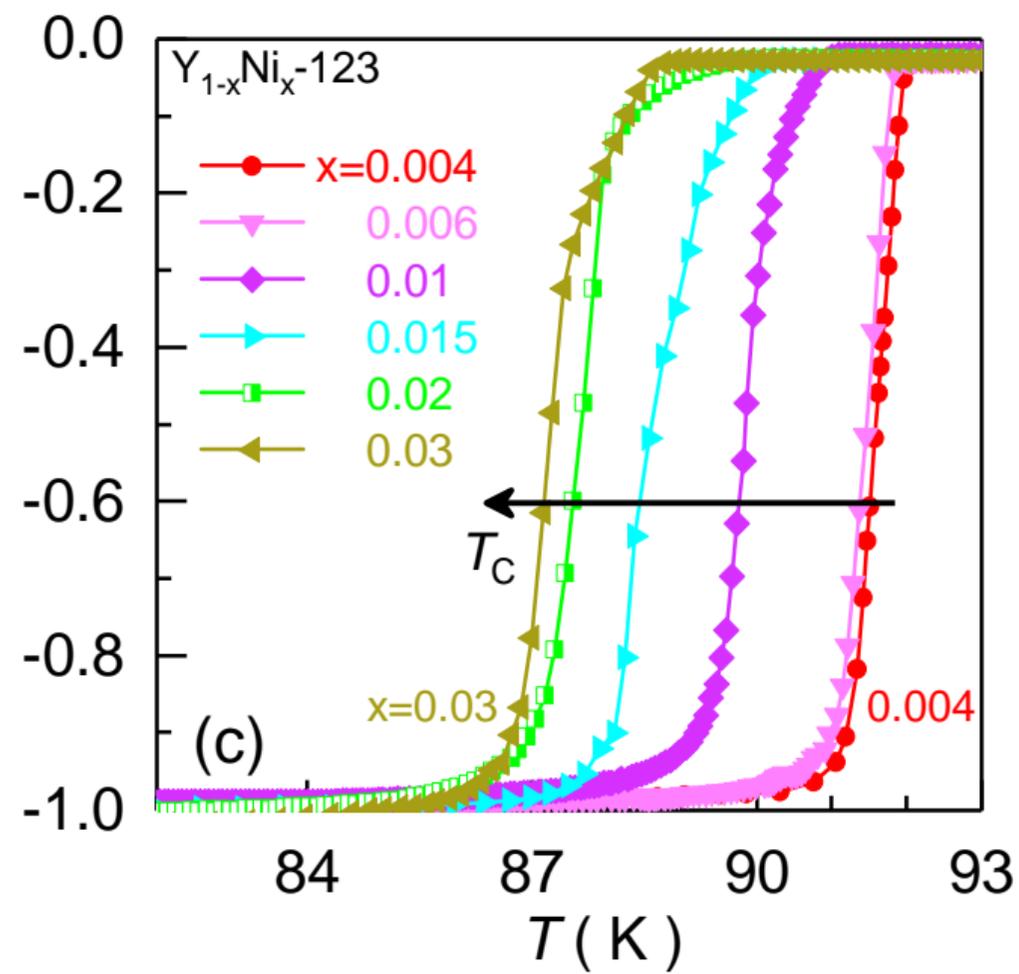

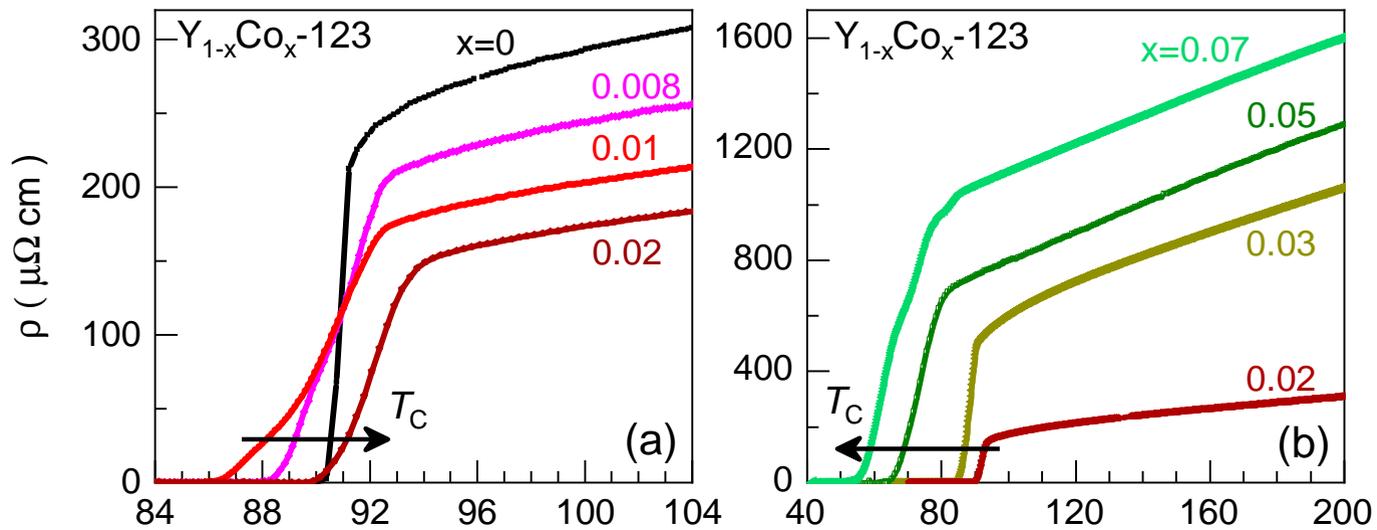
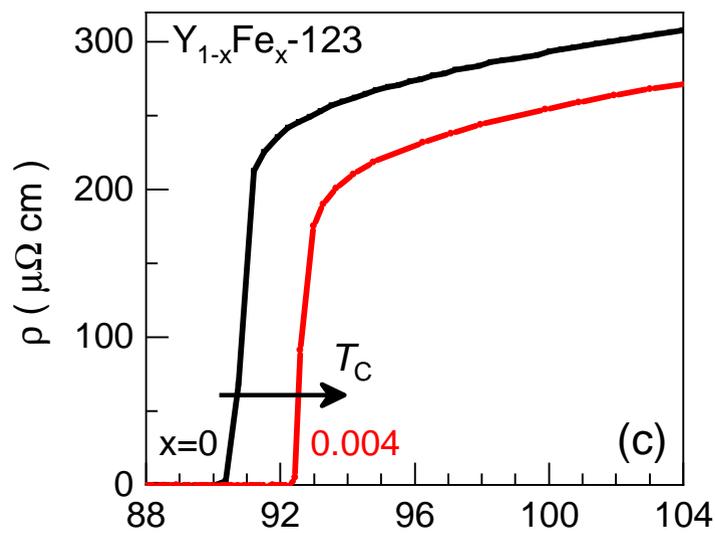
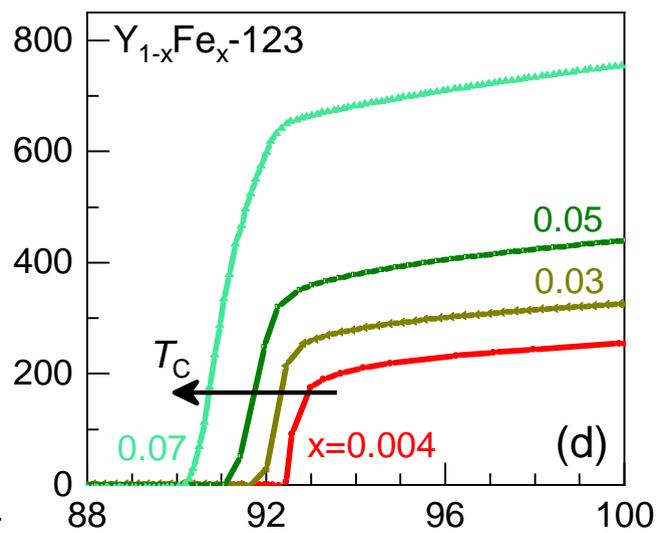
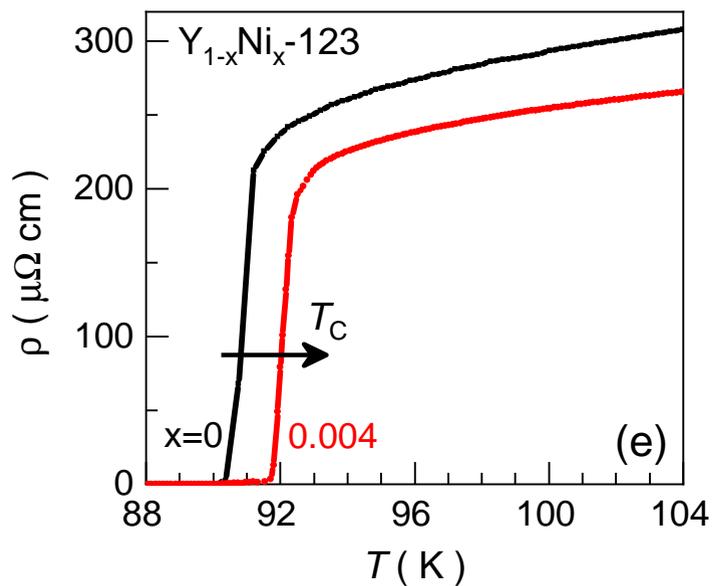
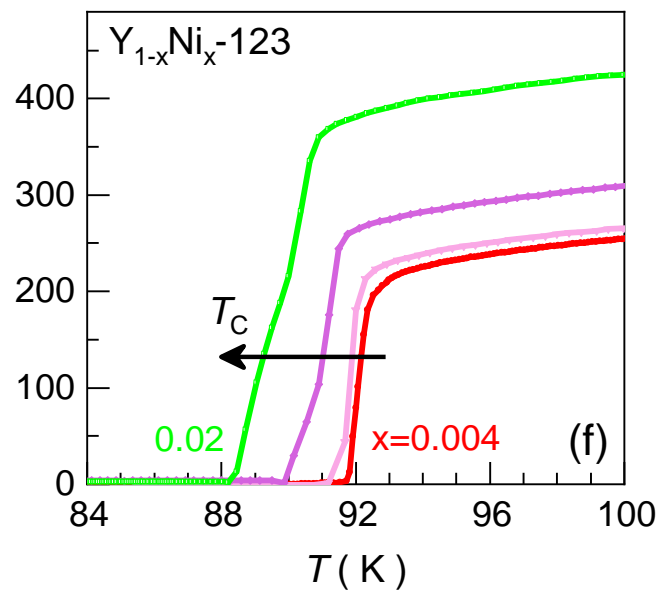

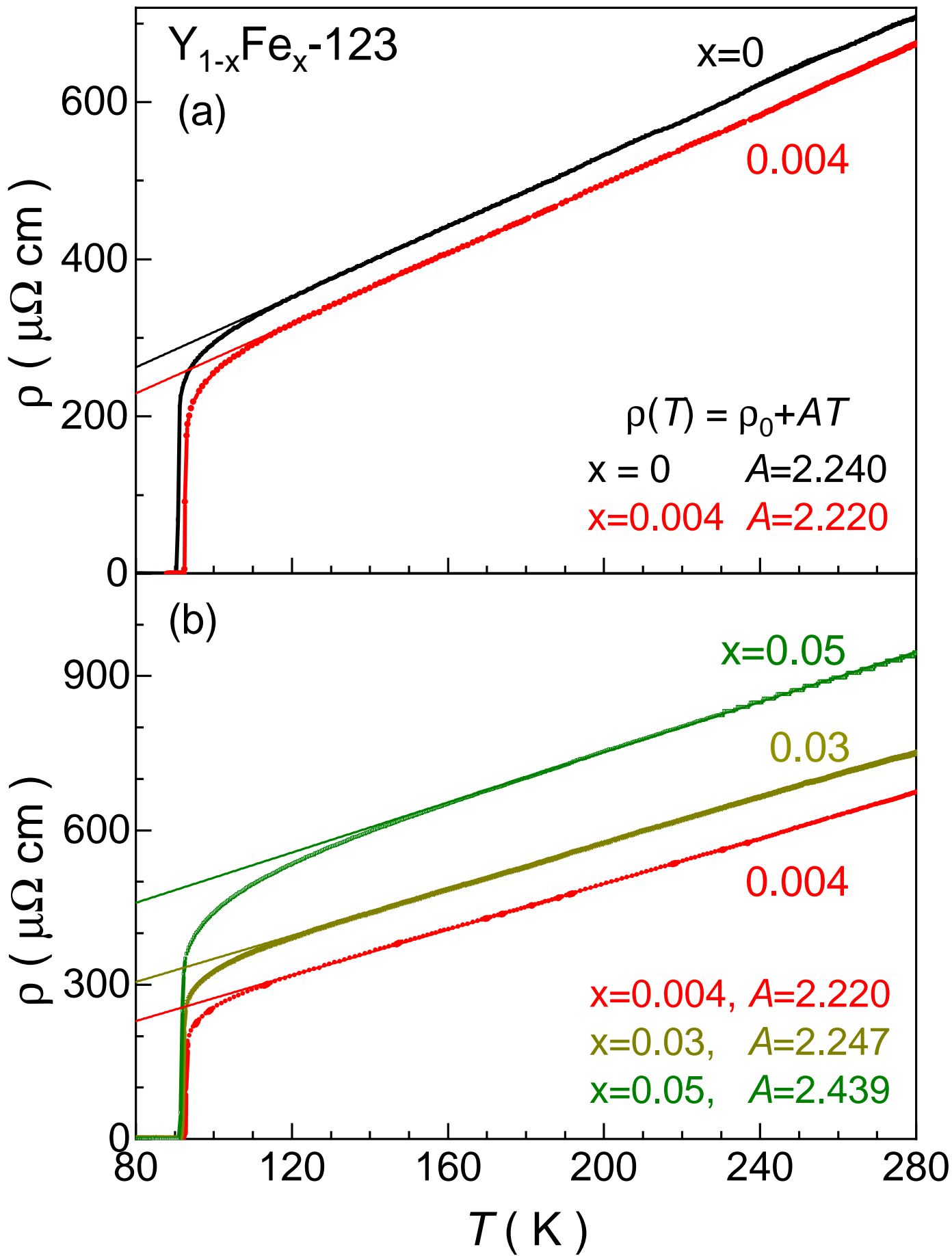

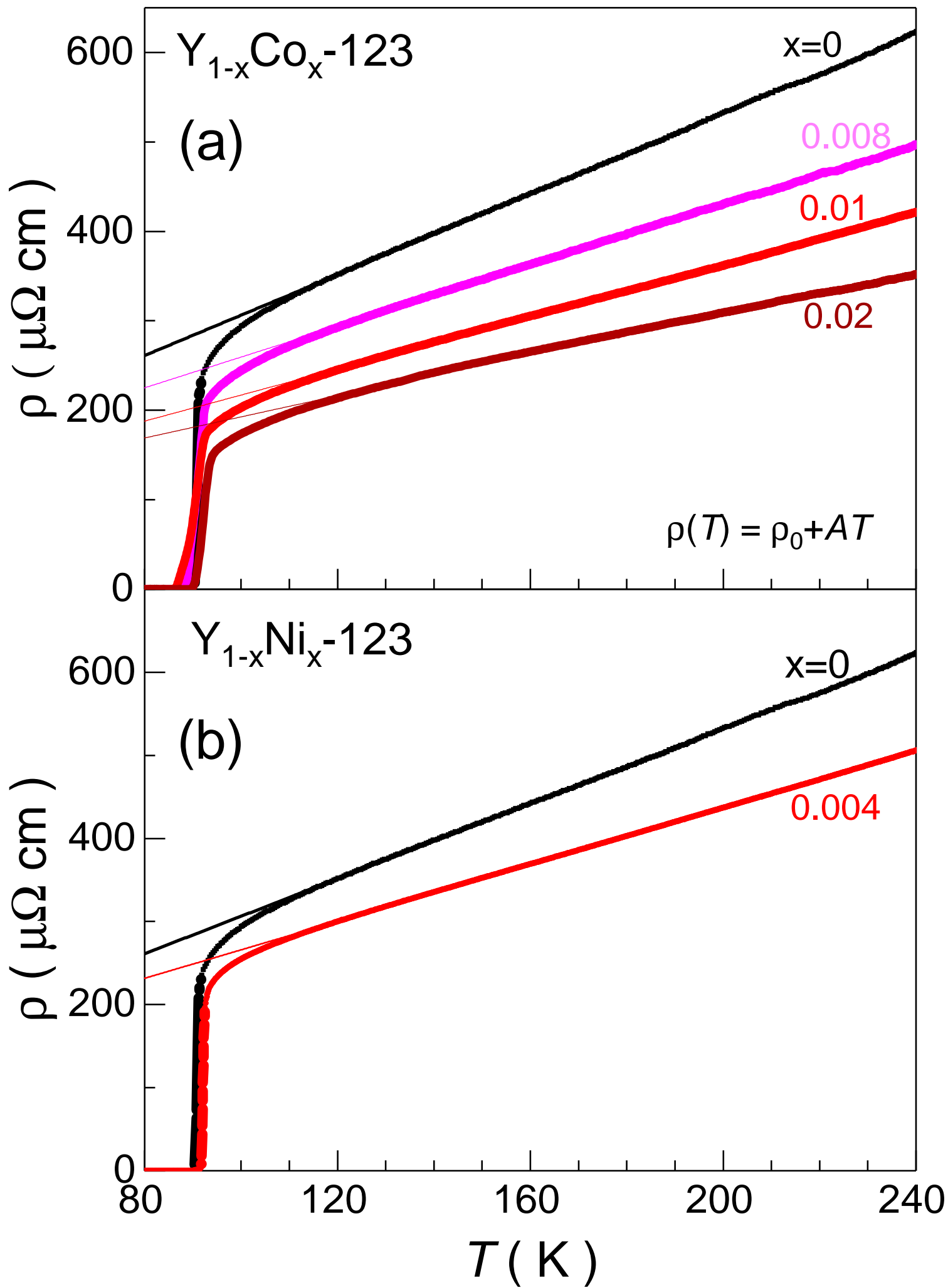

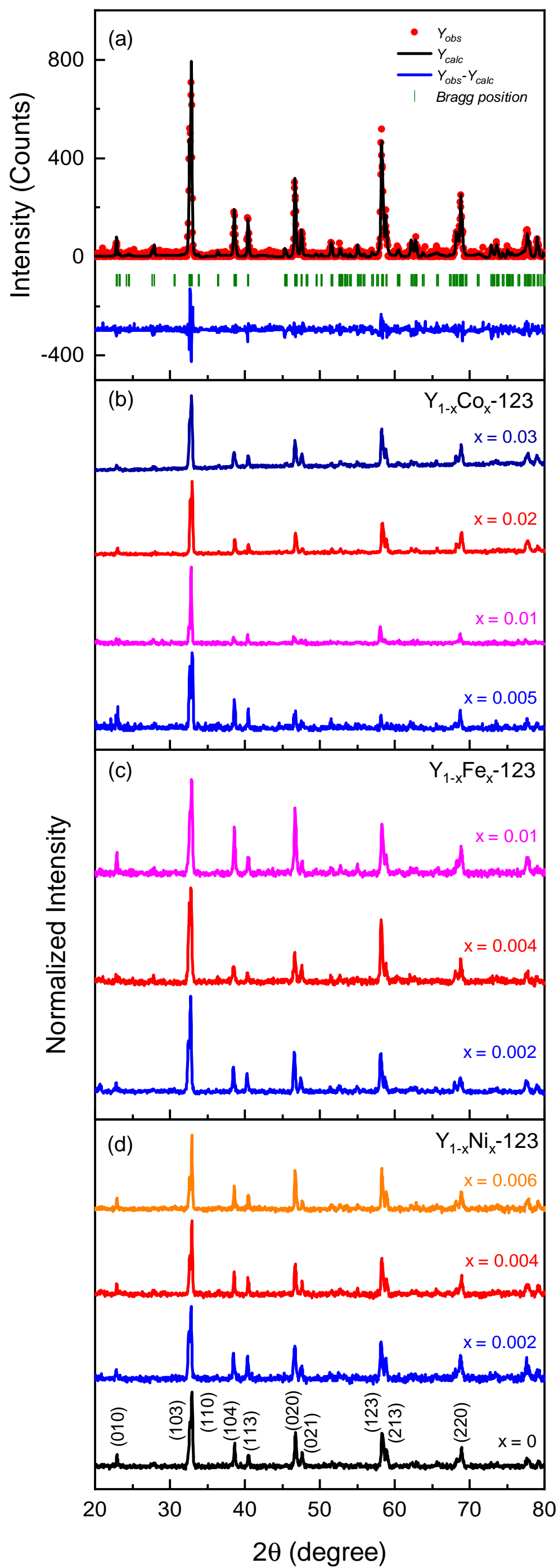

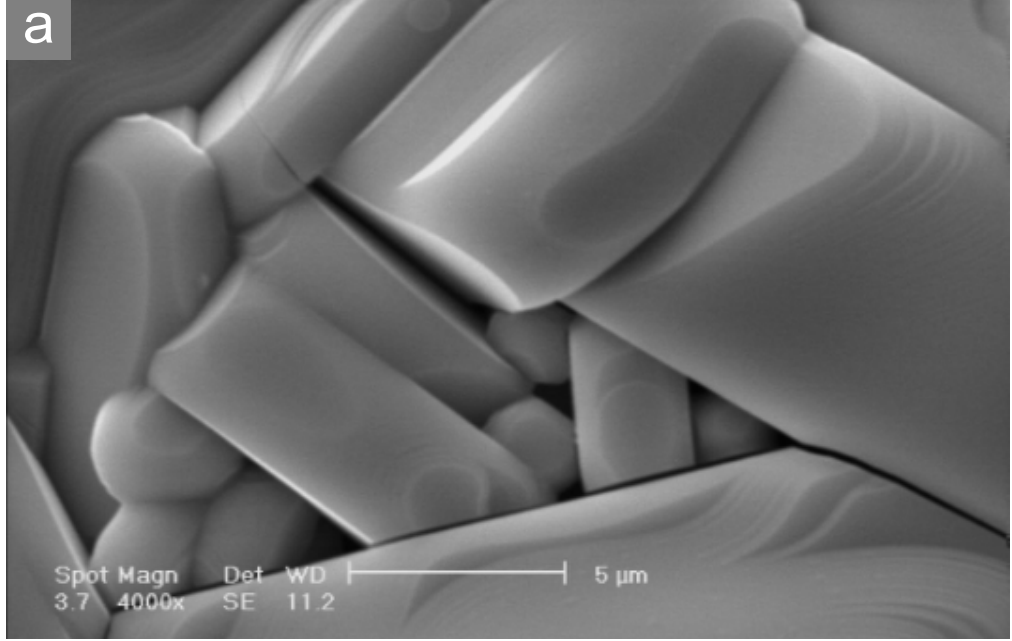
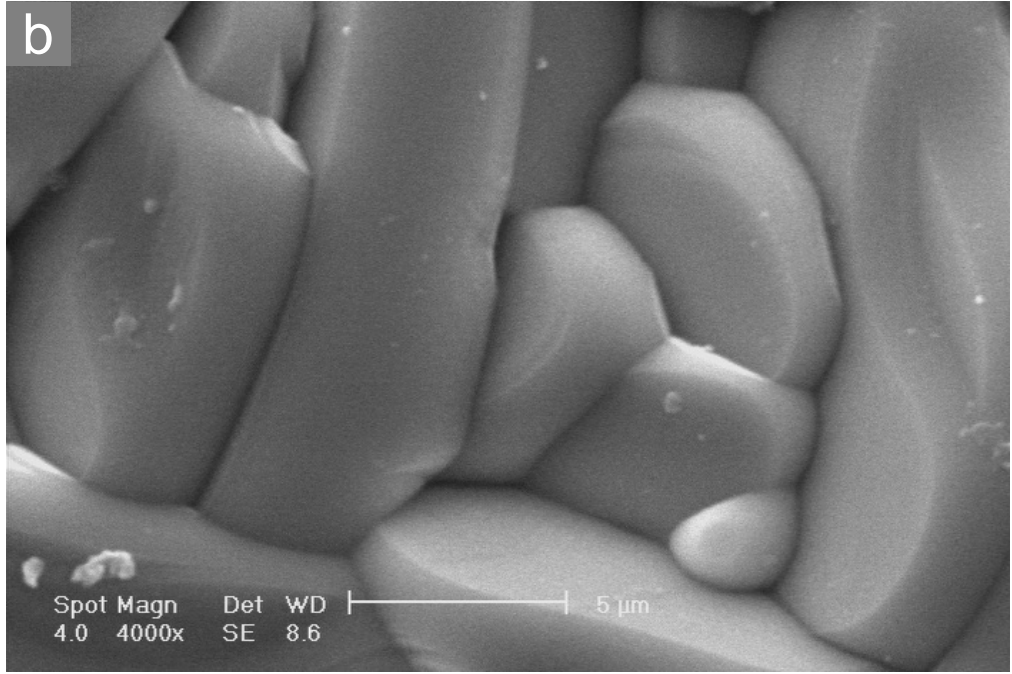
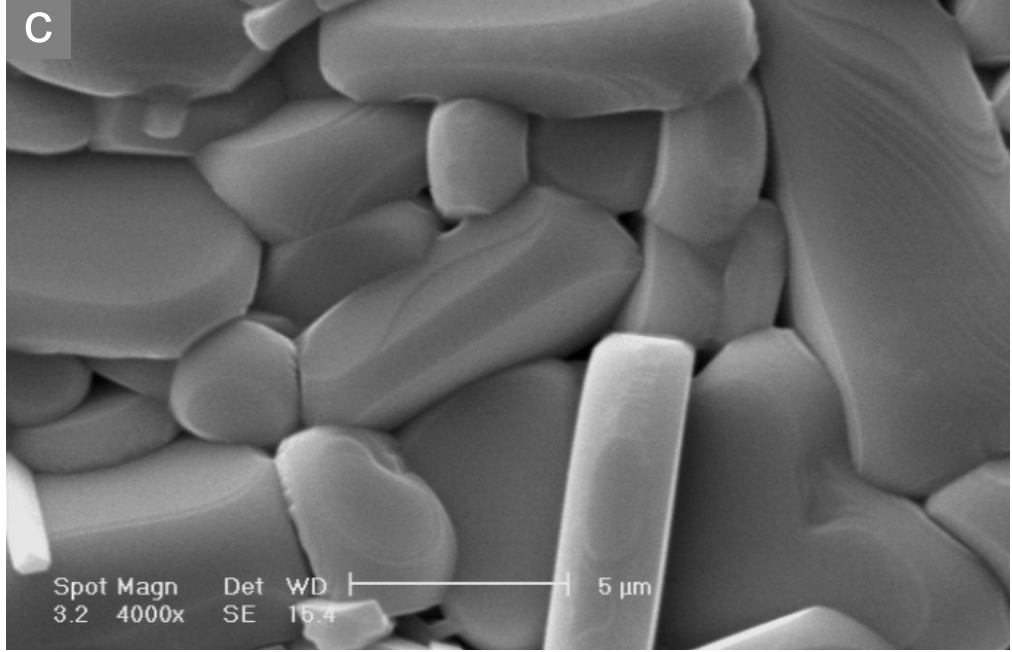